\documentclass[prb,twocolumn,nofootinbib,showpacs]{revtex4}
\usepackage{graphicx}
\usepackage{epsfig}
\usepackage{dcolumn}
\usepackage{bm}
\usepackage{amsmath}
\usepackage{amsbsy}
\usepackage{amssymb}

\newcommand{\bra}[1]{\ensuremath{\langle{#1}|\,}}
\newcommand{\ket}[1]{\ensuremath{\,|{#1}\rangle}}

\begin{document}


\title{Electronic transitions in disc-shaped quantum dots induced
by twisted light}

\author{G.\ F.\ Quinteiro and P.\ I.\ Tamborenea}

\affiliation{Departamento de F\'{\i}sica ``´´Juan Jos\'e
Giambiagi'', Universidad de Buenos Aires, Ciudad Universitaria,
Pabell\'on I, 1428 Ciudad de Buenos Aires, Argentina}

\date{\today}

\begin{abstract}
We theoretically investigate the absorption and emission of light
carrying orbital angular momentum (twisted-light) by
quasi-two-dimensional (disc-shaped) quantum dots in the presence of
a static magnetic field. We calculate the transition matrix element
for the light-matter interaction and use it to explore different
scenarios, depending on the initial and final state of the electron
undergoing the optically-induced transition. We make explicit the
selection rule for the conservation of the z-projection of the
orbital angular momentum. For a realistic set of parameters (quantum
dots size, beam waist, photon energy, etc.) the strength of the
transition induced by twisted light is $10\%$ of that induced by
plane-waves. Finally, our analysis indicates that it may be possible
to select precisely the electronic level one wishes to populate
using the appropriate combination of light-beam parameters
suggesting technological applications to the quantum control of
electronic states in quantum dots.
\end{abstract}

\maketitle



\section{Introduction}

It is a common mathematical procedure to describe light as a
superposition of plane waves, which is possible and convenient
thanks to the linearity of Maxwell's equations and
Fourier analysis.
However, plane waves are only one of the possible \textit{representations}
of light, and despite its wide applicability, other representations may
be better suited to tackle particular problems.\cite{Enk_94}
For example, the expansion of the field in multipolar waves is
normally used to study the radiation associated with nuclei and atoms.
Another, less-known, representation is the so-called twisted light (TL).
As its name suggests, it has an helicoidal wavefront that is
mathematically introduced by a phase with azimuthal dependence,
i.e.\ $e^{-i\,l\phi}$, and a radial dependence of the
\textit{Laguerre-Gaussian} (LG) or \textit{Bessel mode} type.
A peculiarity of the TL representation is that all modes with same
value of $l$ have the same z-projection of orbital angular momentum
(OAM).

The recent development of techniques to generate coherent TL has
boosted the study of many aspects of this peculiar type of radiation.
Research involving TL has been done in numerous areas, such as
generation of twisted beams,\cite{Pad_04} interaction of OAM beams
with mesoscopic particles (optical
tweezers),\cite{Bar02-All02-Fri98} entanglement between spin and OAM
for potential applications to quantum-information
science,\cite{Muthukrishnan02} interaction of OAM fields with atoms
or molecules,\cite{Dav02-Ara05} and QED in confined
geometries,\cite{alw-bab} with twisted beams.
Nevertheless, the problem of the interaction of TL with
condensed matter systems is largely unexplored;\cite{Sim_07}
very recently we have presented the first study on the action of TL
on a bulk semiconductor.\cite{Qui_08}

Quantum dots (QDs) are well-studied man-made nanostructures.\cite{Rei_02,Jac98}
Different types of QDs, such as those produced by lateral confinement
in a 2D electron gas, vertical o stacked, nanowhisker-based, and
self-assembled QDs, are available experimentally.
They all share the basic feature of producing particle confinement in
all directions, which gives rise to a discrete energy spectrum akin
to that of atoms.
Self-assembled, vertical, and nanowhisker QDs are fabricated using
semiconductor materials, and they can confine both electrons and holes.
As the name suggests, disc-shaped QDs (DSQDs) have a stronger
confinement along a given axis, and therefore the dynamics of
electrons and/or holes inside the dot can be restricted to the
quantized in-plane motion.\cite{Bru_92-Tar_96-Ree_88-Pet_01}
In semiconductor-based QDs, transitions between different electronic
states can be accomplished by optical excitation.
The understanding of the optical response of QDs is appealing from a
basic science point of view, but also because of the many possible
applications, for example, to QD lasers, quantum information processing,
and photodetectors.
In spite of the fact that the common Hermite-Gaussian laser pulses are
in wide use to excite QDs,\cite{Haw_85} to the best of our knowledge,
no work has been reported on the interaction of these nanostructures
with twisted light.

Here we develop the theory of optical electronic transitions induced
by TL in semiconductor-based DSQDs.
We find the corresponding selection rule and discuss the use of TL
as a novel and versatile scheme to induce transitions that are not
possible using common optical techniques.
The article is organized as follows.
In Section~\ref{Sec_Model} we review the description of the TL
electromagnetic field and of the electronic states of semiconductor-based
DSQDs in the presence of an external magnetic field.
Section~\ref{Sec_OTrans} is devoted to the calculation of the
matrix elements of the interaction Hamiltonian that couples the light
field to the electrons in the DSQD;
its applications to different cases are treated in
Section~\ref{Sec_Cases}---the special properties of the optical
transitions induced by twisted light in QDs and possible
technological applications become clear in this Section.
Finally, we summarize the results in Section ~\ref{Sec_Conc}.

\section{model}
\label{Sec_Model}

\begin{figure}
  \centerline{\includegraphics[scale=1]{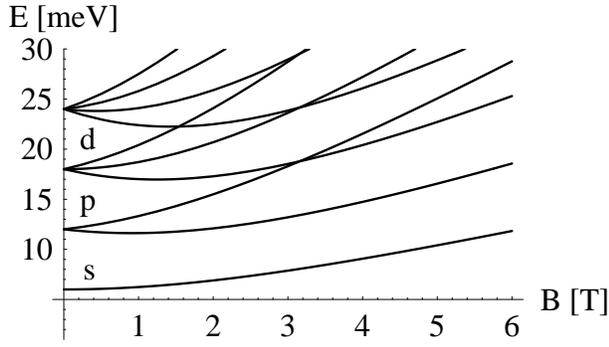}}
  \caption{Energy spectrum as a function of the magnetic field B for
  $\hbar \omega_0 = 6\,$meV and $\hbar \omega_c = 1.7 B\,$meV. The OAM
  of the envelope function is shown using the convention: $n=0(s)$,
  $n=1(p)$, $n=2(d)$, etc.
  \label{fig_levels}}
\end{figure}

For disc-shaped quantum dots, the electron's wave function can be written
as the product of a microscopic cell-periodic function, an envelope
function, and a spin part
$\xi$, that is
\begin{eqnarray}
\label{Eq_WF}
    \psi({\bf{r}})
&=& \left[\phi(r,\theta)\,Z(z)\right] u({\bf{r}}) \,\xi\,.
\end{eqnarray}
For the microscopic function $u({\bf{r}})$, the strain in the QD
lifts the degeneracy of the heavy-hole and light-hole bands; then,
it is safe to restrict the study to a two-band model in the
effective-mass approximation with a conduction and a heavy-hole
bands. For the envelope function
$\left[\phi(r,\theta)\,Z(z)\right]$, the disk shape of the QD allows
for the separation into the vertical ($z$) and the in-plane
$(r,\theta)$ motion.
Furthermore, since the confinement in the $z$-direction is much stronger
than that in the x-y plane, it is a good approximation to assume that
the electron remains in the lowest-energy $z$-eigenstate.
For DSQDs ranging in diameter from tens to a few hundreds of
nanometers,\cite{Bru_92-Tar_96-Ree_88-Pet_01, Bay95} the in-plane
confinement potential can be well approximated by a two-dimensional
harmonic oscillator potential $V_i(r)=\frac{1}{2} m_i^*
\omega_{0i}^2 r^2$, where $i=c,v$ denotes the conduction and valence
bands and $m_i^*$ is the effective electron mass.
The corresponding single-particle problem with the inclusion of an
external magnetic field $B$ applied in the $z$-axis is analytically
solvable yielding eigenfunctions~\cite{Gra_99}
\begin{eqnarray} \label{Eq_EnvWF}
    \phi_{i\,s \,n}(r,\theta)
&=& \frac{(-1)^s}{\sqrt{2\,\pi}\,\ell_i}
    \,\sqrt{\frac{s!}{(s+|n|)!}}
    \,e^{-\frac{r^2}{4\,\ell_i^2}}\,\nonumber\\
&&
    \left( \frac{r}{\sqrt{2}\,\ell_i} \right)^{|n|}
    L_s^{|n|} \left(\frac{r^2}{2\,\ell_i\,^2}\right)
    e^{-i\, n\, \theta} \nonumber\\
&&  = R_{i\,s \,n}(r)\, e^{-i\, n\, \theta} \,,
\end{eqnarray}
where $\ell_i\,^2= \hbar / (2\, |m_i^*| \,\omega_i)$ is a
characteristic length of the confinement of the electrons,
$\omega_i^2=\omega_{0i}^2+\Omega_{i}^2/4$, the cyclotron frequency
$\Omega_{i}=eB/(m_i^*c)$, and $L_s^{|n|}$ is a generalized Laguerre
polynomial with radial quantum number $s$ and $z$-projection of the
OAM $n$.
The energy spectrum of electrons in the conduction band is $E_c =
\hbar \omega_c(2\,s + |n| + 1)-(1/2)\hbar \Omega_{c} \, n$, while
for electrons in the valence band is $E_v = (\hbar \omega_{0c} -E_G)
- [\hbar \omega_v(2\,s + |n|)-(1/2)\hbar \Omega_{v}\, n ]$.
In the presence of an external static magnetic field pointing in the
$z$-direction, the orbital degeneracy of the states is lifted, as shown in
Fig.~\ref{fig_levels}, and for strong enough fields the levels group
into Landau levels (not shown in the figure).

Two different modes of TL beams are nowadays experimentally realizable,
namely, the \textit{Laguerre-Gaussian} (LG) and the \textit{Bessel} modes.
We proceed to develop our theory for a general case, and in the final
stages of the calculation we work with the LG modes in order to take
advantage of the mathematical structure of the electronic wave functions
given in Eq.\ (\ref{Eq_EnvWF}), which contain the generalized Laguerre polynomials.
In the paraxial approximation, the vector potential of the TL beam is
\begin{eqnarray}
\label{EQ_Vpotential}
    \textbf{A}(\textbf{r},t)
&=&  \boldsymbol{\epsilon} \, F(\textbf{r})
    \, e^{i(kz-\omega t)} + c.c. \nonumber \\
&=& \textbf{A}^{(+)}(\textbf{r},t) + \textbf{A}^{(-)}(\textbf{r},t)
\end{eqnarray}
with $\boldsymbol{\epsilon}$ the circular polarization vector normal
to $z$, and for the Laguerre-Gaussian modes
\begin{eqnarray}\label{Eq_F}
    F(\textbf{r})\!\!
&=& \!\!\left[ \frac{C_p^{|l|}}{w_0} \left(
    \frac{\sqrt{2}\,r }{w_0} \right)^{|l|}
    e^{-\frac{r^2}{w_0^2}} L_p^{|l|}
    \left(\frac{2\,r^2}{w_0^2}\right) \right]
    \, e^{-i \,l \,\theta} \nonumber \\
&=& \!\!F_r(r)\, e^{-i \,l \,\theta}
\end{eqnarray}
where $w_0$ is the beam waist, $l$ is the z-projection of the
orbital angular momentum, and $C_p^{|l|}$ is a normalization
constant.~\cite{Rom_02}

Our system, consisting of a single DSQD plus a TL light mode, is
investigated using the semiclassical model where the light is treated
classically and the electrons quantum-mechanically.
For the electron-light interaction we use the minimal-coupling
Hamiltonian and calculate its optical transition matrix elements using
the eigenfuctions given in Eqs.\ (\ref{Eq_WF}) and (\ref{Eq_EnvWF}).
Since our main goal here is to explore the changes in the optical-transition
selection rules brought about by the use of twisted light instead of
plane waves, it is enough to work with a single-particle formalism.
The influence of Coulomb electron-electron interaction, which causes
excitonic effects, modification of the effective confining potential, etc.,
is left for future study.


\section{TL optical dipole matrix elements}
\label{Sec_OTrans}

In order to determine the optical response, we calculate the matrix
element of the transition between single-particle states,
Eq.~(\ref{Eq_WF}), and the light-matter Hamiltonian.
We use the minimal-coupling interaction and retain only the lowest order
in the vector potential.~\footnote{The vector potential in the LG mode is
\textit{not} expressed in the Coulomb gauge; then, the minimal
coupling Hamiltonian $1/(2\,m)(\textbf{p}-q\,\textbf{A})^2$ yields
an extra term:$-(\textbf{p}\cdot\textbf{A})$ where it is understood
that the momentum operator only acts upon the vector potential.
This term need not be considered in our analysis, since it does not
contribute to interband transitions.}
This matrix element is the essential ingredient in many calculations,
e.g.\ Fermi's Golden Rule for the rate of absorption/emission.
To simplify the notation, we use as collective indices the Hebrew
characters Aleph ($\aleph$) and Gimel ($\gimel$) to replace the set
$\{s, n, \xi\}$ [see Eqs.~(\ref{Eq_WF}) and ~(\ref{Eq_EnvWF})],
while the band index will still appear explicitly.
The transition matrix element from an initial state
$\{j, \gimel\}$ to a final state $\{i,\aleph\}$ is
\begin{eqnarray*}
\bra{i \, \aleph} H_{I} \ket{j \, \gimel} &=& - \frac{q}{m}
        \bra{\psi_{i \aleph}} \textbf{A}(\textbf{r}) \cdot \textbf{p}
        \,\ket{\psi_{j \gimel}} \\
    &=& i\hbar  \frac{q}{m} \int_{L^3} d^3r \,
        \psi_{i \aleph}^*(\textbf{r})
        \textbf{A}(\textbf{r})\cdot \nabla
        \psi_{j \gimel}(\textbf{r}) \,.
\end{eqnarray*}
where $q$ is the charge ($-e$) and $m$ is the bare electron mass.
The operation $\nabla \psi_{j\gimel}(\textbf{r})$ yields three
terms and we write $\bra{i \,\aleph} H_{I} \ket{j \,\gimel} = I_1 + I_2 + I_3$
with
\begin{eqnarray}
    I_1 &=& i\hbar \frac{q}{m} \int_{L^3} d^3r \,
        u_{i}({\bf{r}})^*\, u_{j}({\bf{r}})
        \,\textbf{A}(\textbf{r})\cdot \, \nonumber\\
    &&  \left[
        \phi_{i\aleph}({\bf{r}})^*\, \nabla
        \phi_{j\gimel}({\bf{r}})\right]\,\left|Z(z) \right|^2
        \xi_{\aleph}^*\,\xi_{\gimel}  \nonumber\\
    I_2 &=& i\hbar \frac{q}{m} \int_{L^3} d^3r \,
        \phi_{i\aleph}({\bf{r}})^*\, \phi_{j\gimel}({\bf{r}})\,
        u_{i}({\bf{r}})^*\, u_{j}({\bf{r}})
        \,\nonumber\\
    &&  \textbf{A}(\textbf{r})\cdot
        \left[Z(z)^* \nabla Z(z) \right]
        \xi_{\aleph}^*\,\xi_{\gimel} \nonumber\\
    I_3 &=& i\hbar \frac{q}{m} \int_{L^3} d^3r \,
        \phi_{i\aleph}({\bf{r}})^*\, \phi_{j\gimel}({\bf{r}})\,\nonumber\\
    &&  \textbf{A}(\textbf{r})\cdot \left[u_{i}({\bf{r}})^*\,
        \nabla u_{j}({\bf{r}})\right]
        \,\left|Z(z) \right|^2 \xi_{\aleph}^*\,\xi_{\gimel}
        \,.
\end{eqnarray}
The envelope functions and the vector potential are slowly varying
and can be considered constant over a unit cell (lattice constant
$a$); in contrast, the microscopic function $u(\textbf{r})$ is
periodic over unit cells.
These two facts allow for the separation
of the all-space integral ($\int_{L^3}$) into an intra-cell integral
($\int_{a^3}$) and an inter-cell sum ($\sum_{c}$).
Due to the orthogonality of the microscopic function in a cell,
integrals $I_1$ and $I_2$ are nonzero only for intraband transitions
(notice also that $I_2$ represents inter $z$-band transitions, which are
unlikely under strong confinement in the $z$ direction).
As we are interested in optical-frequency transitions which
correspond to interband transitions, we focus our attention on
integral $I_3$
\begin{eqnarray}\label{Eq_I3}
\bra{i \, \aleph} H_{I} \ket{j \, \gimel}
    &=& \frac{-q}{m} \,\xi_{\aleph}^*\,\xi_{\gimel}
        \left[\int_{a^3} d^3r \, u_{i}({\bf{r}})^*\,
        (-i\hbar \nabla) u_{j}({\bf{r}})\right]\cdot
        \nonumber\\
    &&  \hspace{-10mm}\left[\sum_c\, \phi_{i\aleph}({\bf{r}}_c)^*\,
        \phi_{j\gimel}({\bf{r}}_c)\,\left| Z(z_c) \right|^2
        \,\textbf{A}(\textbf{r}_c)\right]\,
        \,,
\end{eqnarray}
which is simplified by taking the continuum limit, thus transforming the
sum over cells to an inter-cell integral according to $\sum
\rightarrow (1/a^3)\int$, and defining the matrix element $a^3\,
{\bf{p}}_{i j} = \int_{a^3} d^3r \, u_{i}({\bf{r}})^*
\,(-i\hbar\nabla) u_{j}({\bf{r}})$.
The vector potential [Eq.~(\ref{EQ_Vpotential})] consists of two
terms.
Inserting the positive part $\textbf{A}^{(+)}(\textbf{r},t)$ in
Eq.~(\ref{Eq_I3}),
\begin{eqnarray}
\bra{i \, \aleph} H_{I}^{(+)} \ket{j \, \gimel}
= -i \, e^{-i \omega
    t}\, \frac{2\,\pi\,\hbar \,q}{m}
    \,(\boldsymbol{\epsilon} \,\cdot \, {\bf{p}}_{i j})
    \,
    \delta_{l,(n_{\aleph}-n_{\gimel})}
    \nonumber \\
  \delta_{\xi_{\aleph},\xi_{\gimel}}\,
    \int_{0}^\infty dr \,r \, F_r(r) \,R_{i\aleph}(r)^*\,
    R_{j\gimel}(r)
    \,,
\end{eqnarray}
where $l$ corresponds to the vector potential and $n$ to the
wave functions, and we assumed the light's wavelength to be much
larger than the height of the QD, so that $e^{i kz} \simeq 1$.

The matrix element $\bra{i \, \aleph} H_{I}^{(+)} \ket{j \, \gimel}$
contains two
terms which represent valence-to-conduction and
conduction-to-valence bands transitions, respectively.
We eliminate one of them by using the rotating-wave approximation (RWA)
and obtain
\begin{eqnarray}\label{Eq_H(+)}
\bra{c \, \aleph} H_{I}^{(+)} \ket{v \, \gimel}
&=& -i \, e^{-i \omega t}\, \frac{2\,\pi\,\hbar \,q}{m}
    \,(\boldsymbol{\epsilon} \,\cdot \, {\bf{p}}_{c v})
    \, \delta_{l,(n_{\aleph}-n_{\gimel})}
    \nonumber \\
  &&\hspace{-10mm}\delta_{\xi_{\aleph},\xi_{\gimel}}\,
    \int_{0}^\infty dr \,r \, F_r(r) \,R_{c\aleph}(r)^*\,
    R_{v\gimel}(r)
    \,,
\end{eqnarray}
for the absorption of light.
The same can be done with
$\textbf{A}^{(-)}(\textbf{r},t)$ yielding
\begin{eqnarray}\label{Eq_H(-)}
\bra{v \, \aleph} H_{I}^{(-)} \ket{c \, \gimel}
&=& -i \, e^{i \omega t}\, \frac{2\,\pi\,\hbar \,q}{m}
    \,(\boldsymbol{\epsilon}^* \,\cdot \, {\bf{p}}_{v c})
    \, \delta_{l,(n_{\gimel}-n_{\aleph})}
    \nonumber \\
  &&\hspace{-10mm}\delta_{\xi_{\aleph},\xi_{\gimel}}\,
    \int_{0}^\infty dr \,r \, F_r(r) \,R_{v\aleph}(r)^*\,
    R_{v\gimel}(r)
    \,,
\end{eqnarray}
for the emission of light.
Further simplification and analysis of Eqs.~(\ref{Eq_H(+)}) and
(\ref{Eq_H(-)}) is only possible once specific functions $F_r(r)$,
$R_{v\aleph}(r)$, and $R_{v\gimel}(r)$ are given. This will be done
in the following section.

Note that the selection rule for the conservation of the z-component
of the OAM in the system of electron plus light field appears
explicitly in Eqs.~(\ref{Eq_H(+)}) and (\ref{Eq_H(-)}).
For absorption and emission processes we get
$\delta_{l, (n_{\aleph}-n_{\gimel})}$ and
$\delta_{l, (n_{\gimel}-n_{\aleph})}$, respectively.


\section{TL-induced optical transitions}
\label{Sec_Cases}

The theory developed in the preceding section will be used here to
study specific cases of optical transitions between QD levels in order
to get a firmer insight into the possibilities opened by excitation
with twisted light.
An important goal is to determine how the beam parameters $\{p, l\}$
[see Eq.\ (\ref{Eq_F})] enable us to choose the allowed transitions
and their strength.
In what follows we will study the excitation process
for the case of Laguerre-Gaussian beams.
Then the matrix element of $H_I^{(-)}$ between initial and final state
yields zero.
From Eq.~(\ref{Eq_H(+)}) with $\gimel=(s\,n\,\alpha)$
and $\aleph=(t \, m \,\beta)$, after inserting the expressions for
$R(r)$ and $F_r(r)$, rearranging terms, and transforming coordinates
to $x=r^2/(2\,\ell_c^2)$, we obtain
\begin{widetext}
\begin{eqnarray}
\bra{c \, \aleph} H_{I}^{(+)} \ket{v \, \gimel}
&=& -i \, e^{-i \omega t}\,
    \frac{2\,\pi\,\hbar \,q}{m}
    \,(\boldsymbol{\epsilon} \,\cdot \, {\bf{p}}_{cv})
    \,\delta_{\alpha,\beta}\,\delta_{l,(m-n)}
    \,\frac{C_p^{|l|}}{w_0}
    \,\frac{(-1)^{s+t}}{2 \pi}
    \sqrt{\frac{t! s!}{(t+|m|)! (s+|n|)!}}
    \nonumber \\
  &&\zeta^{|l|/2}
    \,\int_{0}^\infty dx
    \,x^{(|n|+|m|+|l|)/2}
    \,e^{-x(1+\zeta/2)}
    \,L_p^{|l|} \left(\zeta x\right)
    \,L_t^{|m|} \left(x\right)
    \,L_s^{|n|} \left(x\right),
\label{eq:H+LG}
\end{eqnarray}
\end{widetext}
where we assumed that $\ell_c=\ell_v$~\cite{Ray04} and defined
$\zeta=4\,\ell_c^2/w_0^2$.

Working with Eq.\ (\ref{eq:H+LG}), in Sec.\ \ref{subsec:uppermost}
we analyze the optical transitions induced by twisted light from the
uppermost valence-band state ($s = n = 0$) to a general state in the
conduction band (see Fig.\ \ref{fig_Trans}).
This case offers the mathematical advantage of having only two
generalized Laguerre polynomials in the integral of
Eq.\ (\ref{eq:H+LG}), enabling the use of their orthogonality
relations in order to go farther with the analytical treatment
of the matrix elements.
We concentrate on the physically relevant case of small QDs (small
compared to the size of the beam's waist), in which the parabolic
approximation for the confinement potential is best suited.
For small dots, we compare the strength of the transitions induced by
twisted light with that of the transitions induced by plane waves.
This is done by considering the particular case of a beam without OAM
($l=0$ in Eq.\ (\ref{eq:H+LG})), which resembles plane-wave light.
In Sec.\ \ref{subsec:general} we extend the analysis by solving numerically
Eq.\ (\ref{eq:H+LG}) for a general valence-band initial state and for any
relative sizes of QD and beam waist.
Finally, in Sec.\ \ref{subsec:manipulation} we discuss some of the
possibilities opened by the use of TL in optical excitation experiments
from the point of view of quantum control of the electronic states in QDs.


\subsection{Transitions from uppermost valence-band state}
\label{subsec:uppermost}

Let us consider a QD in its ground state, i.e. with all electrons occupying the
valence-band levels.
The transition of an electron from the valence-band uppermost state to an
arbitrary unoccupied state in the conduction band will be shown to be possible
by choosing the appropriate beam parameters.
The electron in its initial state has wavefunction
$\psi_{v\,0 \,0 \,\beta}({\bf{r}}) = R_{v\,0
\,0}(r)\,\beta$ with $R_{v\,0 \,0}(r) = (\sqrt{2\,\pi}\,\ell_v)^{-1}
\,\exp\left[-r^2/(4\,\ell_v^2)\right]$, while the final excited
state is $\psi_{c\,s \,n \,\alpha}({\bf{r}})$, see Eq.~(\ref{Eq_EnvWF}).
From Eq.~(\ref{eq:H+LG}) with $\aleph=(s\,n\,\alpha)$
and $\gimel=(0\,0\,\beta)$ we arrive at
\begin{eqnarray}\label{Eq_GtoE}
 \bra{c \, \aleph} H_{I}^{(+)} \ket{v \, \gimel}
=  -i \,\frac{C_p^{|l|}}{w_0}
    \, (-1)^s
    \,\sqrt{\frac{s!}{(s+|l|)!}} \nonumber \\
 e^{-i \omega t}\, \frac{\hbar \,q}{m}
    \,(\boldsymbol{\epsilon} \,\cdot \, {\bf{p}}_{cv})
    \,\delta_{\alpha,\beta}\,\delta_{l,n} \, h(\zeta),
\end{eqnarray}
where the dimensionless function $h(\zeta)$ is
\begin{equation}
h(\zeta) = \zeta^{|l|/2} \, \int_{0}^\infty dx  \,x^{|l|}\,
            e^{-x\left(1+\zeta/2\right)}
            \, L_p^{|l|}
            \left(\zeta\,x\right) \,  L_s^{|l|} \left(x\right)\,.
\label{Eq_I}
\end{equation}
For QDs with sizes ranging from $10 \, \text{nm} - 200 \, \text{nm}$,
together with the minimum size of the beam waist $w_0 = 500 \, \text{nm}$,
we obtain $0.001 <\zeta< 0.6$.
Thus, it is reasonable to keep in Eq.~(\ref{Eq_I}) only
the lowest orders in $\zeta$.
Without loss of generality, we assume $l \geq 0$.
We simplify the integral in Eq.\ (\ref{Eq_I}) with the help of
Eq.~(\ref{Eq_Niuk}) to reduce $ L_p^{l}\left(\zeta\,x\right)$.
Thus, we write $h(\zeta) = \zeta^{l/2} \, (I_0 + I_1 + \ldots)$,
with
\begin{eqnarray}
\label{Eq_I0I1}
    I_0
&=& \frac{(l+p)!}{p!}\,\delta_{0 s} \,, \nonumber \\
    I_1
&=& \zeta \,\frac{(l+p)!}{p!}\,\left( p + \frac{l+1}{2} \right)\,
    \left( \delta_{1 s}- \delta_{0 s} \right)    \,,
\end{eqnarray}
where the integral $\int_{0}^\infty dx  \,x^{l}\,e^{-x}\,x\, L_0^l
(x) \,L_s^{l} (x) $ was reduced by using Eq.~(\ref{Eq_Niuk2}) and
the fact that $L_0^l(x)=1$ so $x\, L_0^l(x) \,L_s^{l}(x)= [(1+l)
L_0^l(x)- L_1^{l}(x) ]L_s^{l}(x)$.


%

\medskip

\begin{figure}
  \centerline{\includegraphics[scale=.4]{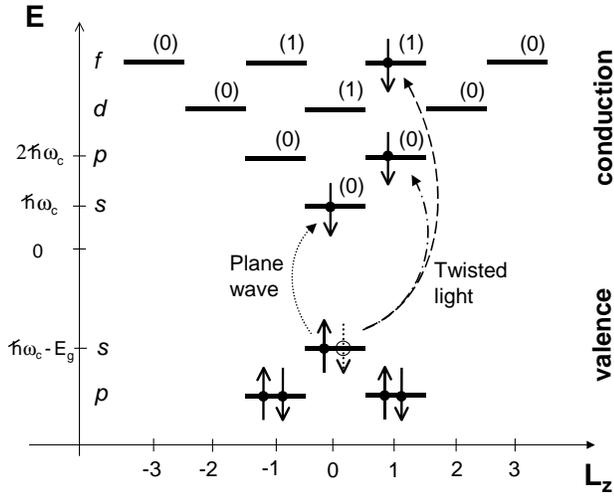}}
  \caption{Schematic representation of the single-particle levels and
  optical transitions for zero magnetic field:
  an electron with spin down has been promoted
  from a valence-band state to a conduction-band state.
  The transition induced by plane waves (dotted line) can only be ``vertical''
  while transitions induced by twisted light (dash-dotted and dashed lines)
  need not be vertical but must
  obey the selection rule for the z-component of OAM;
  we show the transition for light
  carrying OAM $l=1$.
  The convention of Eq.~(\ref{Eq_EnvWF}) is adopted for electronic
  states: $L_z = n$ and the number between parenthesis is
  the radial quantum number $s$. The shells are given by the letters
  $\{s, p, d,\ldots\}$ as customary
  in atomic physics.  \label{fig_Trans}}
\end{figure}

Let us now verify that a ``twisted light'' beam without OAM
($l=0$) yields the same result as plane-wave light.
%
For small QDs, we obtain from Eqs.\ (\ref{Eq_I0I1})
$I_0+I_1=(1-\zeta/2)\delta_{0 j} + \zeta/2 \delta_{1 j}$,
but since an ideal plane wave has $\zeta \rightarrow 0$ we keep only
the zeroth-order term in $\zeta$ in this expression.
If we focus on the lowest-energy transition
$ (v\,0 \,0 \,\alpha) \Rightarrow (c\,0 \,0 \,\alpha)$
we obtain
\begin{eqnarray}
    H_{I}^{(+)}|_{PW}
&\simeq& -i \, e^{-i \omega t}\,
    \frac{C_0^0}{w_0}\, \frac{\hbar \,q}{m}
    \,(\boldsymbol{\epsilon} \,\cdot \, {\bf{p}}_{c v})
    \,;
\end{eqnarray}
this transition, which we named here PW, is depicted in Fig.~\ref{fig_Trans}
as a dotted line labeled ``Plane wave''.
The coefficient $C_0^0 / w_0$ is the amplitude of the vector potential.
Note that the plane-wave limit is obtained by taking
$w_0 \rightarrow \infty$
and
$C_0^0 \rightarrow \infty$ simultaneously, keeping the ratio
constant.

%

We can now compare the relative strength of transitions induced by the usual
PW light and twisted light for small QDs.
In Fig.~\ref{fig_Trans}, TL transitions are shown as dash-dotted and dashed
lines labeled ``Twisted light'', for the particular value $l=1$.
%
%
Retaining the lowest order in $\zeta$ in Eqs.\ (\ref{Eq_I0I1}) also for the
TL mode, we obtain the ratio of amplitudes
\begin{eqnarray}
    \left|\frac{H_{I}^{(+)}|_{TL}}{H_{I}^{(+)}|_{PW}}\right|
=   \frac{C_p^{l}}{C_0^{0}} \,\sqrt{\frac{s!}{(s+l)!}}
    \,\frac{(l+p)!}{p!}  \, \zeta^{l/2}\,.
\label{eq:comp_TL_PW}
\end{eqnarray}
Notice the power-law dependence on $\zeta$, with exponent $l/2$.
For small QDs (i.e.\ small $\zeta$), the transition amplitude with TL becomes
weaker in comparison to that of PW light as the OAM of the light beam
increases.
We expect Eq.\ (\ref{eq:comp_TL_PW}) to be helpful in future experiments on
optical transitions with TL, since it is cast as a comparison with the
standard optical transitions.

\begin{figure}
  \centerline{\includegraphics[scale=1]{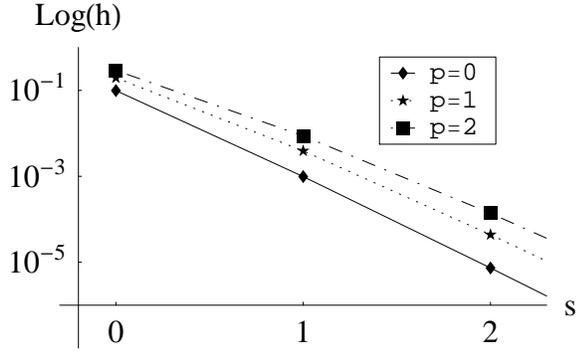}}
  \caption{
  Strength of the optical transition from uppermost valence-band state:
  logarithm of the dimensionless function $h(\zeta)$
  (Eq.~(\ref{Eq_I})) as a function of the radial quantum number $s$ of
  the final electronic state, for $l=1$ and $\zeta=0.01$.
  For this small value of $\zeta$ the TL-induced excitation
  is dominated by the transition from the state $\gimel=(0\,0\,\beta)$
  in the valence band to the state $\aleph=(s\,2\,\beta)$
  with $s=0$ in the conduction band;
  the transition amplitudes to states with $s \neq 0$ are several orders
  of magnitude smaller.
  \label{fig:h_uppermost_strengths}}
\end{figure}

Lastly, let us examine the role of the radial quantum number of the
final state.
%
%
Still considering small dots, Eqs.~(\ref{Eq_I0I1}) suggest that
transitions to all values of the final radial quantum number $s$ are
allowed, and have amplitudes of order $\zeta^{s+|l|/2}$ [shown in
Eqs.\ (\ref{Eq_I0I1}) up to order one].
Fig.\ \ref{fig:h_uppermost_strengths} displays the strength, given by
the function $h(\zeta)$ of Eq.\ (\ref{Eq_I}), of the transitions to
final states with $s=0, 1, 2$, calculated numerically to all orders
in $\zeta$ for $l=1$.
Two of these transitions are illustrated in Fig.~\ref{fig_Trans},
the zeroth-order one shown as a dash-dotted line and the first-order one
as a dashed line.
%
%
In Fig.\ \ref{fig:h_uppermost_strengths} we choose $\zeta= 0.01$.
For this small value of $\zeta$ the TL-induced excitation is dominated by
the transition from the valence-band state $\gimel=(0\,0\,\beta)$ to the
conduction-band state $\aleph=(s\,2\,\beta)$ with $s=0$.
There is a strong dependence on $s$, and it can be seen that the transition
amplitudes to states with $s \neq 0$ are several orders of magnitude smaller.
We will see in the next section that this situation changes
radically for QDs and beam waists of comparable sizes.


\subsection{Transitions from general valence-band state}
\label{subsec:general}

Here we consider the transition of an electron initially in a
general valence-band state $\psi_{v \gimel}({\bf{r}})$
to a conduction-band state $\psi_{c \aleph}({\bf{r}})$,
with collective indices $\gimel = (s \,n \, \alpha)$ and
$\aleph = (t\, m \, \beta)$.
The analytical solution of the integral appearing in
Eq.~(\ref{eq:H+LG}), although possible, is cumbersome (see
Appendix~\ref{Ap_Any}).
Therefore, we present numerical results for the function
\begin{eqnarray}
h(\zeta) =
    \zeta^{|l|/2}
    \,\int_{0}^\infty dx
    \,x^{(|n|+|m|+|l|)/2}
    \,e^{-x(1+\zeta/2)} \nonumber \\
    \,L_p^{|l|} \left(\zeta x\right)
    \,L_t^{|m|} \left(x\right)
    \,L_s^{|n|} \left(x\right)
\label{eq:h_general}
\end{eqnarray}
which allow us to explore all values of $\zeta$.
A value of $\zeta > 1$ represents a situation of a narrow
beam interacting with a large structure.
Even though the parabolic approximation for the confining potential
is likely not to be valid for quantum disks of such sizes, our
calculation may provide qualitatively correct results.

\begin{widetext}
\begin{figure}
  \centerline{\includegraphics[scale=1]{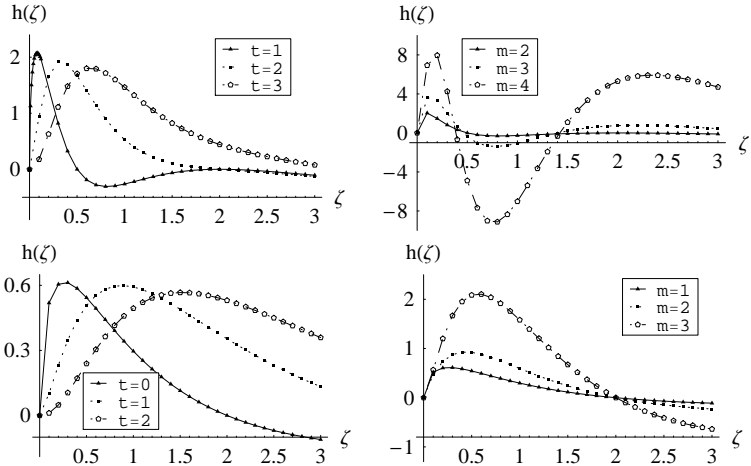}}
  \caption{Strength of the optical transition from general valence-band
  states: dimensionless function $h(\zeta)$ (Eq.~(\ref{eq:h_general})).
  Light-parameter $p=1$ in all cases.
  Top left: from state ($s=1, n=1$) to ($t, m=2$).
  Top right: from state ($s=1, n=1$) to ($t=1, m$).
  Bottom left: from state ($s=0, n=0$) to ($t, m=1$).
  Bottom right: from state ($s=0, n=0$) to ($t=0, m$).
  Transitions from the same initial state are displayed on the same row.
  Curves on the left column represent transitions to final states
  with varying radial quantum number $t$.
  Curves on the right column represent transitions to final states
  with varying z-projection of the OAM $m$.}
  \label{fig:general_cases}
\end{figure}
\end{widetext}

In Fig.~\ref{fig:general_cases} we plot $h(\zeta)$ from
Eq.\ (\ref{eq:h_general}).
We consider two different initial states.
The bottom row corresponds to transitions from the uppermost
valence-band state, and it is thus an extension of the results
of the previous section to general values of $\zeta$.
On the bottom-left panel we plot three possible transitions
to final states differing only in their radial quantum number $t$.
Pictorially, these states would lie, in Fig.\ \ref{fig_Trans}, on a vertical
line.
We see that for values of $\zeta$ up to about 0.5 (small QDs) the
transition to the final state with $s=0$ is the dominant one, which
is consistent with the analytical results of the previous section
for $\zeta \simeq 0$.
With increasing $\zeta$ the relative strength of the different
transitions is altered, and for $\zeta > 1.4$ the order of the three
transitions is inverted compared to the situation at $\zeta < 0.5$.
The bottom-right panel shows transitions to final states with
varying z-projection of the OAM, for fixed radial quantum number
$t=0$.
These transitions can be visualized in Fig.\ \ref{fig_Trans}
as having final states lying on the lowest diagonal going up and
to the right.
The top row of Fig.~\ref{fig:general_cases} presents analogous
results but for transitions originating in the valence-band state
$(s=1, n=1)$, chosen somewhat arbitrarily to illustrate the more
general case. Besides a change of scale from top to bottom panels,
the similarities among them are obvious.

The relative strength of transitions induced by plane waves and those
induced by TL can be considered again, now for transitions with
an arbitrary initial state.
We cannot say much for the case of large values of $\zeta$, beyond the
fact that a dependence on $\zeta$ will be present, according to
Eq.\ (\ref{eq:h_general}).
However, for $\zeta \rightarrow 0$ we note that the integral
in $h(\zeta)$ tends to a constant (see Appendix \ref{Ap_Any}) and we recover
the result of $h(\zeta) \propto \zeta^{|l|/2}$
obtained in the previous section.


\subsection{Manipulation of electronic states}
\label{subsec:manipulation}

Our study suggests that twisted light is a versatile tool to optically
manipulate states in QDs.
By a smart choice of beam parameters, QD size, and external static magnetic
field one can select precisely the electronic level to be populated.
In the following, we illustrate, with two examples, how this can be
accomplished.

Let us assume that we wish to connect the uppermost valence-band
state with the $f$-shell conduction-band state $(n=1,s=1)$ [see Eq.\
(\ref{Eq_EnvWF})] on a small QD ($\zeta\ll 1$);
this transition is depicted as a dashed line in Fig.\ \ref{fig_Trans}.
We first choose the light-beam parameters: tune the laser on
resonance (or close) with the energy difference between these states
($\Delta E= E_g + 3\hbar\omega_c$) and take $l=1$, $p=0$ (from our
previous analysis, the value of $p$ is not very important).
With this requirements, not only the desired transition is possible,
but also another one that promotes an electron from the $p$-shell
valence band state $(n=-1,s=0)$ to the $d$-shell conduction band
state $(n=1,s=0)$.
Nevertheless, if a static magnetic field is applied, one of these
transitions can be moved off resonance from the light (see in
Sect.\ \ref{Sec_Model} the general expressions for the energy levels with
magnetic field), leaving only one dominant transition.

For our second example we will assume that the laser cannot be tuned
precisely to match the energy difference between a couple of states,
and that we deal with a QD of larger size, say about $400 \, \text{nm}$.
This last assumption means that, by modifying the light-beam waist,
we can choose $\zeta$ on the interval $(0,1)$.
Then, by selecting the light-beam parameter $l$ we instantly eliminate
transitions between states differing more than $l$ in their z-projection
of OAM.
Furthermore, by adjusting the beam-waist, we see that we can change
the probability of transition depending on the value of the radial
quantum number of the electronic states (see Fig.\
\ref{fig:general_cases}).
To fix ideas, let us imagine that the emission line of the light beam is
centered around $E_g + 2\hbar\omega_c$ with width $\hbar\omega_c$.
Under these conditions, transitions from the uppermost valence-band
state to shells $p,d,f$ in the conduction-band are possible.
If we choose beam parameters $l=1$ and $p=0$, and waist such that
$\zeta \simeq 1$ the dominant transition will become that to the
$f$-shell conduction-band state $(t=1,m=1)$.
Thus, we conclude that even if the laser cannot be precisely tuned, we
still can decide which will be the dominant transitions.


\section{Concluding remarks}
\label{Sec_Conc}

We have investigated the theory of optical absorption of twisted
light---carrying orbital angular momentum---by disk-shaped
semiconductor-based quantum dots in the presence of a static
magnetic field, and calculated the dipole transition matrix element,
which is the basic building block for other specific studies of optical
response.

As a first general result we made explicit the selection rule for the
conservation of the z-projection of the orbital angular momentum.
In addition, we observe that not only vertical but also more general
transitions are possible, due to the fact that the z-component of
the orbital angular momentum of a TL beam can assume any integer
value.

Different scenarios were explored according to what the initial and
final states of the electron in the QD are.
In the first place, we studied the transitions of an electron from
the uppermost valence-band state to any conduction-band states, and
obtained analytical expressions in terms of powers of the ratio
$\zeta$ of the QD size to the light-beam waist.
For realistic values of $\zeta$ the strength of the transition
induced by TL can be around $10\%$ of the value of the transition
using common laser (non-twisted) fields.
We also considered the transitions that bring an electron from any
valence-band states to any conduction-band states.
This case was studied for larger values of $\zeta$, and it enabled us to
qualitatively think of the physics of similar systems, such
as large QDs or quantum disks.
Finally, we found that $\zeta$ plays also a role in selecting the most
likely transitions.
In all scenarios, we found that a smart choice of beam parameters
(z-projection of OAM, radial quantum number, energy, beam waist)
allows one to select which state will the electron be promoted to.

The current availability of sources of coherent beams of TL permit
us to envisage their use for more versatile quantum control of electronic
states in QDs and the indirect control of magnetization
in doped QDs.
Our theoretical analysis suggests that it may be possible to select
precisely the electronic level one wishes to populate using the
appropriate combination of beam parameters (and nano-structure
size).
In particular, derived formulas and results for $\zeta$ around $0.1$
are directly applicable to disk-shaped semiconductor-based quantum
dots; therefore, our predictions may be experimentally verified
using current technology.

We have left unexplored the interesting question of the local
nanometric-scale magnetic field generation.
It should be clear that---although this effect may be small---the absorption
of twisted-light photons entails a transfer of OAM to the system with
the resulting emergence of an electronic current within the QD,
which in turn produces a magnetic field.~\cite{Qui_08}


\begin{acknowledgments}

We are grateful to Pawel Hawrylak for useful discussions.
We acknowledge support through grants ANPCyT PICT-2006-02134,
CONICET PIP 5851, and UBACyT X495.
G.F.Q. also acknowledges support from the National Research Council
of Canada (Ottawa).
P.I.T.\ is a researcher of CONICET.

\end{acknowledgments}


\vspace*{0.3cm}

\appendix
\section{Laguerre polynomials} \label{App_Laguerre}

The Laguerre polynomials verify the following identities. The
orthogonality condition
\begin{equation}\label{Eq_Laguerre}
    \int_0^{\infty} e^{-x}\,x^\alpha\,L_n^{\alpha}(x) \,L_m^{\alpha}(x)
        dx = \frac{\Gamma(n+\alpha+1)}{n!} \, \delta_{nm} \,;
\end{equation}
the relation
\begin{eqnarray}\label{Eq_Niuk}
    L_m^\beta (\tau\,x)
&=& \sum_{n=0}^\infty \left(\begin{array}{c}
    \beta+m \\  m-n \end{array}\right) \tau^n\,(1-\tau)^{m-n}
    \, L_n^\beta (x) \nonumber\\
&\simeq& \left(\begin{array}{c} \beta+m \\  m
    \end{array}\right) (1 - m\,\tau) \, L_0^\beta (x) + \nonumber\\
&&  \left(\begin{array}{c}
    \beta+m \\  m-1 \end{array}\right) \tau \, L_1^\beta (x)
\end{eqnarray}
from Niukkanen,~\cite{Niu_85} and the second line was deduced for
$\tau \rightarrow 0$. Also,
\begin{eqnarray}\label{Eq_Niuk2}
    x
&=& (1+l) L_0^l(x) - L_1^{l} (x)
\end{eqnarray}
and
\begin{eqnarray}\label{Eq_Niuk3}
    L_m^\beta(x)
&=& \sum_{n} \frac{(\beta-\alpha)_{m-n}}{(m-n)!} L_n^\alpha(x)
\end{eqnarray}
where $(a)_i=\Gamma(a+i)/\Gamma(a)$ is a Pochhammer symbol. Finally
\begin{eqnarray}\label{Eq_Niuk4}
    L_m^k (x)\,L_n^l (x)
&=& \sum_{\alpha=0}^\infty (-1)^\alpha D_\alpha
    L_{m+n-\alpha}^{k+l}(x)\,.~~
\end{eqnarray}


\section{Analytical solution for any-to-any transition}
\label{Ap_Any}

Without loss of generality, we make $l,n,m>0$ in Eq.\
(\ref{eq:h_general})
\begin{eqnarray}
\label{Eq_EtoEHI}
    h(\zeta)
  &=& \zeta^{l/2}
    \int_{0}^\infty dx
    \,x^{m}
    \,e^{-x\left( 1 + \zeta/2 \right)} \nonumber \\
  &&\,L_p^{m-n} \left(\zeta\,x\right)
    \,L_t^{m} \left(x\right)
    \,L_s^{n} \left(x\right)  \,.
\end{eqnarray}
and solve the integral, for small $\zeta$, using Eq.\
(\ref{Eq_Niuk}), Eq.~(\ref{Eq_Niuk3}), and $L_0^{n-m}(x) = 1$.
Separating into orders of $\zeta$
\begin{eqnarray}
    I_0
&=& \left(\begin{array}{c} m-n+p \\  p
    \end{array}\right)
    \frac{(n-m)_{s-t}}{(s-t)!}
    \frac{\Gamma(t+m+1)}{t!}   \nonumber \\
    I_1
&=& \zeta\,\left(\begin{array}{c} l+p \\  p
    \end{array}\right)
    \frac{(n-m)_{s-t}}{(s-t)!}
    \frac{\Gamma(t+m+1)}{t!}
    \left(\frac{1+l}{2}-p\right)  \nonumber \\
&&  +\zeta
    \left[
        \left(\begin{array}{c} l+p \\
        p-1 \end{array}\right) +
        \frac{1}{2}  \left(\begin{array}{c}
        l+p \\  p \end{array}\right)
    \right]    \, I_{11}\,.
\end{eqnarray}
where $(a)_i=\Gamma(a+i)/\Gamma(a)$ is a Pochhammer symbol and
$I_{11} = \int_{0}^\infty dx \,x^{m} \,e^{-x} \,L_1^l (x) \,L_t^{m}
(x) \,L_s^{n} (x)$ can be formally reduced using
Eq.~(\ref{Eq_Niuk4}).

Eq.~(\ref{Eq_EtoEHI}) admits also a simple solution for the
particular case of $p=0$, following the same reasoning as done in
Section\ \ref{subsec:uppermost}.



\newpage


\end{document}